\documentclass[conference]{IEEEtran}
\IEEEoverridecommandlockouts

\usepackage[utf8]{inputenc}
\usepackage[T1]{fontenc}

\usepackage{cite}
\usepackage{amsmath,amssymb,amsfonts}
\usepackage{algorithmic}
\usepackage{graphicx}
\usepackage{textcomp}
\usepackage{xcolor}
\usepackage{tikz}
\usepackage[top=0.75in,left=.75in,right=.75in,bottom=1.05in]{geometry}

\def\BibTeX{{\rm B\kern-.05em{\sc i\kern-.025em b}\kern-.08em
    T\kern-.1667em\lower.7ex\hbox{E}\kern-.125emX}}
\begin{document}

\title{Payment Channels with Proofs
  \thanks{
    The results were supported by the Ministry of Education, Youth and Sports within the dedicated program ERC CZ under the project POSTMAN no. LL1902, Czech Science Foundation grant no. 25-17929X, Amazon Research Awards, and the ERC PoC grant \emph{FormalWeb3} no. 101156734.
  }
}

\author{\IEEEauthorblockN{1\textsuperscript{st} Chad E. Brown}
\IEEEauthorblockA{\textit{Czech Technical University in Prague} \\
Prague, Czech Republic}
\and
\IEEEauthorblockN{2\textsuperscript{nd} Cezary Kaliszyk}
\IEEEauthorblockA{\textit{University of Melbourne} \\
\textit{University of Innsbruck} \\
Melbourne, Australia \\
\texttt{ckaliszyk@unimelb.edu.au}\\
0000-0002-8273-6059}
\and
\IEEEauthorblockN{3\textsuperscript{rd} Josef Urban}
\IEEEauthorblockA{\textit{Czech Technical University in Prague} \\
Prague, Czech Republic \\
\texttt{josef.urban@gmail.com}\\
0000-0002-1384-1613}
}

\maketitle

\begin{abstract}
  The fundamental building blocks of the Bitcoin lightning network are bidirectional payment channels.
  We describe an extension of payment channels in the Proofgold network
  which allow the two parties to bet on whether a proposition will be proven by a certain time.
  These provide the foundation for a Proofgold lightning network that would
  allow parties to request proofs (by betting there will be no proof by a certain time)
  and other parties to provide proofs (and be rewarded by betting there will be a proof).
  The bets may also provide a way to approximate the probability that a certain proposition is
  provable (in the given amount of time). We describe the implementation of
  payment channels supporting proofs in Proofgold and discuss a potential lightning network
  that could be built as a result.
  One application of such lightning network would be a
  large decentralized infrastructure for fast collaborative formalization projects.
\end{abstract}


\section{Introduction}

The Bitcoin lightning network~\cite{LightningNetworkWhitePaper}
provides a Layer 2 solution for securely and quickly sending off-chain
payments, addressing the limitation that on-chain Bitcoin transactions
can take a long time to confirm. Similar constructions could be applied
to other cryptocurrencies where the primary use case is the transfer of
value and there is little or no need for more complex transaction data.

One of the most important use-cases of Proofgold~\cite{Proofgold2022},
is being able to pay users for proofs of theorems. Similar ideas as the
Bitcoin lightning network could be used to create a lightning network
for Proofgold, however, the ideas underlying the Bitcoin
lightning network must be extended to allow users on the network to
also pay for proofs of theorems
by betting on whether a theorem will be proven by a given deadline.
We describe such an extension in this paper.
The bets could also be used as a prediction market, to estimate
the probability that a proof of a proposition will be public
by a certain deadline.

Proofgold is a cryptocurrency supporting formal logic and mathematics
launched in 2020. In Proofgold users can publish theories (primitives and axioms)
within a framework of intuitionistic higher-order logic.
These theories can be developed as users publish documents, making new definitions
and proving new theorems within the theory.
As of early 2025, six theories have been created in Proofgold,
with the most developed one being a higher-order set theory~\cite{BrownPak19}.
In addition, over 2000 definitions and over 50000 proofs have been published
into the Proofgold blockchain.
Proofgold users can also place bounties on conjectures
they would like to see proven.
As of early 2025, over 13000 bounties have been placed on conjectures
and almost 3000 have been collected (as the result of someone proving the
conjecture or its negation).

The usual method of publishing Proofgold documents onto the chain
is time consuming. Proofgold has a block time approximating one block per hour.
Furthermore, a commit-and-reveal scheme is used to protect document
authors from plagiarism. As a consequence, it takes about a day to publish
a document into the Proofgold blockchain.
An appropriate off-chain lightning network would
provide a faster alternative to placing a bounty on a conjecture (on chain)
and waiting for someone to publish a document (on chain)
with a corresponding proof of the conjecture in order to collect the bounty.
Since the lightning network would work off-chain, we will also need an alternative
to the commit-and-reveal scheme Proofgold uses
(on chain) to ensure authorship of proofs.
The alternative we consider is paying agents who bet there will be a proof by
a certain deadline, regardless of who authored the proof.
Those agents will be paid by other agents who bet there will not be a proof
(presumably because they want to pay for a proof to appear).

\subsection*{Contents:} In Section~\ref{s:lightning} we remind the concept
of payment channels and the ideas underlying the Bitcoin lightning network.
We propose the extension of payment channels to proofs in
Section~\ref{s:goldchannel}. In section \ref{s:goldlightning} we discuss an
extension of the channels to a full lightning network for Proofgold and in
section \ref{s:pmarket} we discuss betting on the odds of theorems being
proved that leads to a prediction market for theorems. In Section
\ref{s:impl} we discuss the implementation of the discussed extensions in
the Proofgold Lava client.

\section{Payment Channels and Lightning}\label{s:lightning}

The fundamental building blocks of the lightning network (for Bitcoin)
are bidirectional payment channels
(or, simply ``payment channels'').
We give a simple explanation here, and leave the curious reader to study~\cite{LightningNetworkWhitePaper} for more details.

A payment channel involves two agents, e.g., Alice and Bob.
To open a payment channel, Alice and Bob combine some bitcoin into a single
asset (bitcoin utxo). For example, suppose Alice and Bob both have 1 bitcoin
and they together build a transaction spending each of their bitcoins to
a new address with 2 bitcoins. The new address is a 2-of-2 multisig address,
so it can only be spent if Alice and Bob agree on how to spend it.
(That is, they must {\em{both}} sign any transaction spending from the address.)

Of course, if they simply do this, then Bob can hold Alice's bitcoin hostage
-- only agreeing to spend the 2 bitcoins if he receives, e.g., 1.9 bitcoins
and Alice receives only 0.1. Likewise, Alice can hold Bob's bitcoin hostage
for the same reason.
So, before sending the bitcoins to the multisig address,
Alice and Bob create three unsigned transactions:
a ``funding transaction'' sending Alice and Bob's bitcoin to the combined multisig address
and two initial ``commitment transactions.''
Before Alice or Bob sign and publish the funding transactions,
each must sign one of the commitment transactions. The commitment transactions
reflect the current balance of the payment channel.
We will oversimplify and say both commitment transactions spend from
the (currently hypothetical) utxo from the funding transaction
with two outputs: 1 bitcoin for Alice and 1 bitcoin for Bob.
Alice creates this commitment transaction, signs it, and sends it to Bob.
Bob does the same.
Now both can sign and publish the funding transaction,
so that the 2 bitcoins at the multisig address is an actual asset (utxo) with 2 bitcoins.
These 2 bitcoins cannot be held hostage by either Alice or Bob, since either
can sign and publish their commitment transaction (which was already signed by the other party).

The procedure described above does allow Alice and Bob to create an asset (essentially ``opening a channel'')
and spend the asset to recover their balance (essentially ``closing a channel''),
but lacks the ability to {\emph{update}} the balance.
Without the ability to update the balance, payments are not yet possible.
In order to allow for payments, the commitment transactions use
hash timelock contracts (htlcs).

Described simply, an htlc is a certain kind of script that can be spent in one of two ways:
by one party after enough confirmations or by another party (immediately) with a ``secret.''
Extending the example above, suppose the two initial commitment transactions Alice and Bob did 
are as follows:
\begin{itemize}
\item Alice's initial commitment transaction spends the multisig utxo
  with 1 bitcoin to Bob
  and 1 bitcoin to an htlc address.
  The htlc address is spendable by Alice after 3 days
  but is also spendable by Bob immediately if Bob knows Alice's secret.
  Assume only Alice knows Alice's secret initially.
\item Bob's initial commitment transaction is symmetric, spending the multisig utxo
  with 1 bitcoin to Alice
  and 1 bitcoin to an htlc address.
  The htlc address is spendable by Bob after 3 days
  or by Alice immediately with Bob's secret.
  Again, assume only Bob knows Bob's secret initially.
\end{itemize}
Assume Alice signs Bob's initial commitment transaction and
Bob signs Alice's initial commitment transaction
before they both sign the funding transaction.
After the funding transaction has been confirmed, the payment channel is open.
Alice can close the channel by signing and publishing her initial commitment transaction.
This would give Bob access to his 1 bitcoin immediately and give Alice access to her 1 bitcoin
after 3 days. Assuming Bob does not know Alice's secret, her 1 bitcoin will be unspent and available
after those 3 days.
Likewise Bob could close the channel by signing and publishing his initial commitment transaction,
causing him to wait 3 days for his 1 bitcoin to be available.

The use of htlc's and secrets makes the current pair commitment transactions {\emph{revokable}}.
This is what allows Alice and Bob to update their balance.
For example, suppose Alice wishes to send 0.1 bitcoin to Bob over the channel.
Assuming the two of them agree, they create a new pair of commitment transactions
(with new secrets) sending 0.9 to Alice and 1.1 to Bob.
They then reveal the previous secrets to each other.
Now that Alice and Bob know the previous secrets, neither can close the channel using
the old commitment transactions without losing their entire balance in the channel.
Thus they can still close the channel at any time, but can only do so by using
their latest commitment transaction.

It will be convenient to have a notation for htlc scripts.
Let $h$ be the (SHA256) hash of a secret $s$ (a 256 bit number),
$N$ be a number of blocks and $\alpha$ and $\beta$ be addresses (the hash of
the public key of a private key)
and let
${\mathsf{h}}(h,\alpha,N,\beta)$ denote the htlc script that
can be spent in two ways (where which way is chosen by the spender):
\begin{enumerate}
\item A secret $s$ hashing to $h$ is given and the transaction is signed by $\alpha$. This behavior uses cryptographic operations common to both the Bitcoin and Proofgold scripting languages.
\item At least $N$ blocks have passed and the transaction is signed by $\alpha$.
  Checking if $N$ blocks have passed is enforced by {\tt{OP\_CSV}} (check sequence verify), an operation common to both the Bitcoin and Proofgold scripting languages.
\end{enumerate}
Suppose $\alpha$ is Alice's address and $\beta$ is Bob's address.
In practice the $N$ would be chosen to be a specific number.
For example, in Bitcoin taking $N=288$ would require $288$ confirmations (roughly 2 days)
before ${\mathsf{h}}(h,\alpha,N,\beta)$ is spendable by Bob.
This would give Alice roughly two days to notice if Bob publishes an outdated
commitment transaction.
Each of Alice's commitment transactions sends Alice's balance to an address
controlled by the script ${\mathsf{h}}(h^A_m,\beta,N,\alpha)$ and Bob's balance to $\beta$.
Here $h^A_m$ is the hash of Alice's $m^{th}$ secret $s^A_m$.
Each of Bob's commitment transactions sends Alice's balance to $\alpha$
and Bob's balance to ${\mathsf{h}}(h^B_m,\alpha,N,\beta)$.
Here $h^B_m$ is the hash of Bob's $m^{th}$ secret $s^B_m$.
If $m$ is the most recent pair of commitment transactions, then
both parties should know $s^A_i$ and $s^B_i$ for each $i<m$,
but only Alice should know $s^A_m$ and only Bob should know $s^B_m$.


Once there are multiple agents with open payment channels,
it is clear that one can implement a network of such channels.
As a simple example, suppose Alice and Bob have an open payment channel
and Bob and Charlie have an open payment channel.
Suppose Alice has a balance of at least 1 bitcoin in her payment channel with Bob,
and that Bob has a balance of at least 1 bitcoin in his payment channel with Charlie.
In this case Alice can send Charlie 1 bitcoin by
sending Bob 1 bitcoin via the first payment channel
while Bob (atomically) sends 1 bitcoin to Charlie via the second payment channel.
Note that Alice's balance has decreased by 1 bitcoin and Charlie's balance has increased by 1 bitcoin.
Bob's total balance has not changed: his balance in the first payment channel has increased by
1 bitcoin and his balance in the second payment channel has decreased by 1 bitcoin.
In order to incentivize Bob to play this intermediate role, he may charge Alice a small fee.
For example, Bob could require Alice to send Bob 1 bitcoin and 1 satoshi on the first channel
while Bob still sends only 1 bitcoin to Charlie on the second channel.

The lightning network is essentially a network of payment channels where payments
can be routed through multiple payment channels. There are a number of details (such as
finding routes with sufficient liquidity) which are beyond the scope of this simple explanation.

\section{Payment Channels with Proofs in Proofgold}\label{s:goldchannel}

Since Proofgold supports multisig and htlc addresses, bidirectional payment channels
as described for Bitcoin can also be opened, updated, and closed in Proofgold. 
A lightning network is arguably more important on Proofgold, as Proofgold's Layer 1
is slower than Bitcoin's; Bitcoin has 1 block every 10 minutes on average
while Proofgold has 1 block an hour on average.
Suppose Alice and Bob have an open payment channel
with a 2-of-2 multisig utxo with 200 Proofgold bars
and a current balance of 100 bars for Alice and 100 bars for Bob.
The current balance is reflected by the latest pair of commitment transactions
that Alice and Bob can use to close the channel.

Suppose Alice wants a proof of proposition $P$.
In Proofgold (Layer 1), she could put a bounty on the proposition $P$.
Whoever proves $P$ (or its negation) would be able to collect the bounty.
Proofgold uses a commitment scheme so that the author of a proof
can safely publish a proof without having it ``stolen'' during the publication process.
The author first publishes a commitment and (roughly 12 hours) later publishes
the document with the proof. If someone else wanted to ``steal'' the proof,
they would first see it when the document is being published. They would then
need to publish their own commitment and wait another (roughly) 12 hours before
they could publish their version of the document (essentially replacing the original
author with themselves). Presumably, the original document with the original author
would have been published during this waiting period.
The first author to publish the proof of a proposition $P$
becomes the ``owner'' of $P$. The owner of $P$ can collect any bounties on $P$.

Lifting the bounty mechanism to a Layer 2 lightning network seems impossible.
One cannot use the commitment scheme since that relies on Layer 1,
which is slow. This conflicts with the desire for Layer 2 to be fast
and not need to wait for blocks on Layer 1. Layer 1 should only be used to
open and close channels.

Since Alice and Bob have an open channel, she could simply announce
to Bob that she will pay 50 bars for a proof of $P$.
Suppose Bob proves $P$. If he sends her the proof, she has no reason
to update the balance to reflect paying Bob 50 bars, and Bob has no recourse.
If he claims to have the proof but will only reveal it after Alice sends the
50 bars, she has no recourse if she sends the 50 bars but Bob never reveals a proof.
Ultimately, Layer 1 should be the ``recourse'' if one party is uncooperative.
That is, if one party is uncooperative, the other party should be able to close the
channel in a way that reflects the expected balance.

Suppose Alice and Bob update the balance using new commitment transactions with
three outputs instead of two.
One output would send 50 bars to Alice and the second output would send 100 bars to Bob.
One of these first two outputs would be an htlc address (depending on which
of the pair of commitment transactions it is).
The third output would send 50 bars to a third address.
The intention of this third address is that the 50 bars go to Bob if Bob
has a proof of $P$
and the 50 bars go to Alice otherwise.
Assuming we can do this, then Bob can safely reveal the proof of $P$ to Alice
privately. Alice should then update the balance of the channel to reflect
that she has 50 bars and Bob has 150 bars, without the need for a third output.
Otherwise, Alice is uncooperative and Bob can publish the proof of $P$ on the
blockchain and close the channel by publishing the commitment transaction with the
third output. Since $P$ will have been proven on Layer 1, Bob should be able to
spend the third output with 50 bars to himself.
On the other hand, if Bob tries to close the channel without $P$ being proven,
Alice should be able to collect the third output (after an appropriate timeout,
assuming Alice has requested a proof of $P$ by a certain time).

Note that in this high-level description it is not important that Bob actually
be the one to prove $P$. If anyone else publishes a proof of $P$ on Layer 1,
then Bob can still use this to collect the 50 bars.
As a consequence, the 50 bars can be seen as a ``bet'' by Alice that Bob will
not know a proof of $P$ by some timeout and a ``bet'' by Bob that he
will know a proof of $P$ by the timeout (even if the proof was given by someone else).
With the current setup, however, it is not quite a ``bet'' since Alice makes
no reward if no one provides a proof by Alice's timeout.
Also, since the bet is specific to paying Bob, but not specific to Bob proving $P$,
it is natural to question why Alice would make this bet with Bob instead
of someone else
with whom she has a payment channel.

Bob could encourage Alice to make the bet with him instead of another partner
by adding 10 bars of his own to the bet.
After making such a bet, the commitment transactions
would send 50 bars to Alice, 90 bars to Bob
and 60 bars to the third output.
In this scenario, Alice will be rewarded 10 bars for ``winning'' the bet because no one
proved $P$ in time.
We will now assume this modification, so that the third output is worth 60 bars.

As a first approximation, the third output should be spendable by a script
that lets Bob spend the third output if $P$ has been proven
and lets Alice spend the third output after a certain time has passed.
The key element of Proofgold's scripting language that allows for such
an output is an operation {\tt{OP\_PROVEN}} which checks if a given proposition (identified by its hash root)
has been proven in the Proofgold blockchain.
Let us call such a script a {\em{ptlc (proposition time lock contract)}}
and denote it by ${\mathsf{p}}(P,\beta,T,\alpha)$ for a proposition $P$, an address $\beta$,
a time (either in block height or unix timestamp) $T$ and an address $\alpha$.
In the discussion above, $\beta$ is Bob's address and $\alpha$ is Alice's address.
The script ${\mathsf{p}}(P,\beta,T,\alpha)$ states that the output
can be spent in two different ways (which way is determined by spender):
\begin{enumerate}
\item The proposition $P$ has been proven (on chain) and the transaction is signed by the private key address for $\beta$. This is where {\tt{OP\_PROVEN}} is used, in addition to the usual operations for signature checking.
\item Time $T$ has passed and the transaction is signed by the private key for $\alpha$. This uses the usual operations for signature checking and {\tt{OP\_CLTV}} (check locktime verify). {\tt{OP\_CLTV}} is an operation common to both Bitcoin and Proofgold and checks if a certain time has passed (either judged by block height or by the unix timestamps of blocks).
\end{enumerate}

Using the ptlc ${\mathsf{p}}(P,\beta,T,\alpha)$ does work to enforce the bet as described above, but has the drawback that it is not revokable. Suppose Bob does prove $P$ (off chain)
and Alice acknowledges this
by updating the payment channel to reflect Alice's balance of 50 bars and Bob's balance of
150 bars. Updating the payment channel requires Alice to reveal the secret she used
for the htlc of her previous commitment transaction (with the third output corresponding to the bet).
Alice could, nevertheless, wait until time $T$ has passed and publish her revoked
commitment transaction with the third output. If she does this, Bob can take the first two outputs,
using his signature and Alice's revealed secret. This allows Bob to take 140 bars.
However, he cannot take the third output -- with 60 bars -- until he publishes the
proof of $P$ on chain. Before he does so, Alice can spend the third output (since
time $T$ has passed) and obtain 60 bars. In effect, Alice has been able to use
a revoked commitment transaction to withdraw more than her intended balance.

We can remedy this situation by combining the htlc behavior with the ptlc behavior.
That is, we also need the third output to be controlled by the counterparty
after a secret has been revealed.
We simply denote this as the composition ${\mathsf{h}}(h,\delta,N,{\mathsf{p}}(P,\beta,T,\alpha))$
where $h$ is the hash of a secret,
$N$ is a number of blocks, $P$ is a proposition, $T$ is a time
and $\delta$, $\beta$ and $\alpha$ are addresses.
The script
${\mathsf{h}}(h,\delta,N,{\mathsf{p}}(P,\beta,T,\alpha))$
states that the output
can be spent in two different ways (chosen by the spender):
\begin{enumerate}
\item A secret $s$ hashing to $h$ is given and the transaction is signed by $\delta$.
\item At least $N$ blocks have passed and inputs for the script ${\mathsf{p}}(P,\beta,T,\alpha)$ are given.
\end{enumerate}
Note that there are actually three ways of spending the output, since
the script ${\mathsf{p}}(P,\beta,T,\alpha)$ can be satisfied in two ways.
Note also that though there are three addresses $\alpha$, $\beta$ and $\delta$,
the intention is that $\delta$ is either $\alpha$ or $\beta$ (whichever is the
counterparty of the commitment transaction).

\begin{figure}
  \begin{center}
  \begin{tikzpicture}[
    node distance=1.5cm and 2.5cm,
    every node/.style={font=\small, align=center},
    txn/.style={draw, thick, rounded corners, fill=gray!20, minimum width=3cm, minimum height=1cm},
    output/.style={draw, thick, rounded corners, fill=blue!20, minimum width=3.5cm, minimum height=0.8cm},
    arrow/.style={thick, ->, >=stealth}
    ]

\node[txn] (tx) {Channel fund};

\node[output, xshift=5cm, yshift=0.5cm](out1){100 bars \\ \( \alpha \) };
\node[output, xshift=5cm, yshift=-0.5cm](out2){100 bars \\ \( {\mathsf{h}}(h^B_1,\alpha,48,\beta) \)};

\draw[arrow] (tx.east) -- (out1.west);
\draw[arrow] (tx.east) -- (out2.west);

  \end{tikzpicture}
  \end{center}
  \caption{Bob's initial commitment transaction $\tau^B_1$}\label{fig:bobinitcomm}
\end{figure}
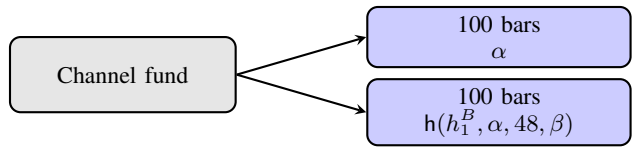

Let us reconsider the example above.
We will assume the value of $N$ is $48$, so that the htlc behavior
requires a user to wait for 48 confirmations (roughly 2 days in Proofgold).
Alice and Bob both make initial commitment transactions before the funding transaction
is signed, transmitted and confirmed.
Alice's initial commitment transaction $\tau^A_1$ will use secret $s^A_1$ with hash $h^A_1$
and will have two outputs: 100 bars to ${\mathsf{h}}(h^A_1,\beta,48,\alpha)$
and 100 bars to $\beta$, as shown in Figure~\ref{fig:aliceinitcomm}.
For Bob's initial commitment transaction $\tau^B_1$, the roles of $\alpha$ and $\beta$ are reversed,
with an output of 100 bars to $\alpha$ and 100 bars to ${\mathsf{h}}(h^B_1,\alpha,48,\beta)$, as shown in Figure~\ref{fig:bobinitcomm}.
After these initial commitment transactions are signed by the counterparty
and the funding transaction has confirmed, the channel is open.

\begin{figure}
  \begin{center}
  \begin{tikzpicture}[
    node distance=1.5cm and 2.5cm,
    every node/.style={font=\small, align=center},
    txn/.style={draw, thick, rounded corners, fill=gray!20, minimum width=3cm, minimum height=1cm},
    output/.style={draw, thick, rounded corners, fill=blue!20, minimum width=3.5cm, minimum height=0.8cm},
    arrow/.style={thick, ->, >=stealth}
    ]

\node[txn] (tx) {Channel fund};

\node[output, xshift=5cm, yshift=0.5cm](out1){100 bars \\ \( {\mathsf{h}}(h^A_1,\beta,48,\alpha) \)};
\node[output, xshift=5cm, yshift=-0.5cm](out2){100 bars \\ \( \beta \) };

\draw[arrow] (tx.east) -- (out1.west);
\draw[arrow] (tx.east) -- (out2.west);

  \end{tikzpicture}
  \end{center}
  \caption{Alice's initial commitment transaction $\tau^A_1$}\label{fig:aliceinitcomm}
\end{figure}
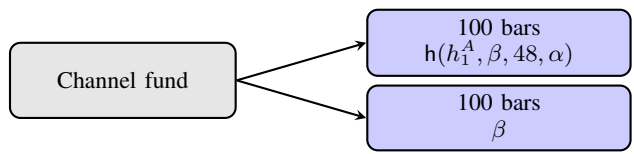

\begin{figure}
  \begin{center}
  \begin{tikzpicture}[
    node distance=1.5cm and 2.5cm,
    every node/.style={font=\small, align=center},
    txn/.style={draw, thick, rounded corners, fill=gray!20, minimum width=3cm, minimum height=1cm},
    output/.style={draw, thick, rounded corners, fill=blue!20, minimum width=3.5cm, minimum height=0.8cm},
    arrow/.style={thick, ->, >=stealth}
    ]

\node[txn] (tx) {Channel fund};

\node[output, xshift=5cm, yshift=1cm](out1){50 bars \\ \( \alpha \) };
\node[output, xshift=5cm, yshift=0cm](out2){90 bars \\ \( {\mathsf{h}}(h^B_2,\alpha,48,\beta) \)};
\node[output, xshift=5cm, yshift=-1cm](out3){60 bars \\ \( {\mathsf{h}}(h^B_2,\alpha,48,{\mathsf{p}}(P,\beta,T+48,\alpha)) \) };

\draw[arrow] (tx.east) -- (out1.west);
\draw[arrow] (tx.east) -- (out2.west);
\draw[arrow] (tx.east) -- (out3.west);

  \end{tikzpicture}
  \end{center}
  \caption{Bob's commitment transaction $\tau^B_2$ with the bet that Bob will have a proof of $P$ before block height $T$}\label{fig:bobbetcomm}
\end{figure}
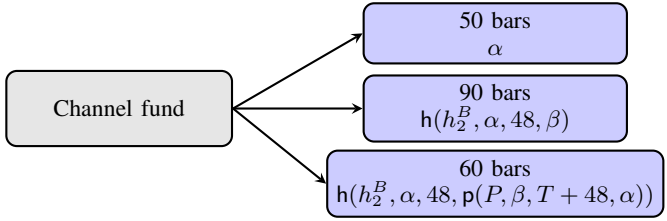

Suppose Alice and Bob agree to bet on whether or not Bob will have (access to) a proof of $P$
at some block height $T$ in the future,
with Alice contributing 50 bars and Bob contributing 10 bars.
Since the htlc behavior will lead to a delay of 48 blocks,
we use the block height $T+48$ in the ptlc below.
Alice's new commitment transaction $\tau^A_2$ will have three outputs:
$50$ bars to ${\mathsf{h}}(h^A_2,\beta,48,\alpha)$,
$90$ bars to $\beta$
and $60$ bars to ${\mathsf{h}}(h^A_2,\beta,48,{\mathsf{p}}(P,\beta,T+48,\alpha))$,
as shown in Figure~\ref{fig:alicebetcomm}.
Bob's new commitment transaction $\tau^B_2$ will have three outputs:
$50$ bars to $\alpha$
$90$ bars to ${\mathsf{h}}(h^B_2,\alpha,48,\beta)$,
and $60$ bars to ${\mathsf{h}}(h^B_2,\alpha,48,{\mathsf{p}}(P,\beta,T+48,\alpha))$,
as shown in Figure~\ref{fig:bobbetcomm}.
After the counterparty has signed the new commitment transactions,
Alice and Bob revoke the previous commitment transactions by sharing the
secrets $s^A_1$ and $s^B_1$.

\begin{figure}
  \vspace{0.1in} 
  \begin{center}
  \begin{tikzpicture}[
    node distance=1.5cm and 2.5cm,
    every node/.style={font=\small, align=center},
    txn/.style={draw, thick, rounded corners, fill=gray!20, minimum width=3cm, minimum height=1cm},
    output/.style={draw, thick, rounded corners, fill=blue!20, minimum width=3.5cm, minimum height=0.8cm},
    arrow/.style={thick, ->, >=stealth}
    ]

\node[txn] (tx) {Channel fund};

\node[output, xshift=5cm, yshift=1cm](out1){50 bars \\ \( {\mathsf{h}}(h^A_2,\beta,48,\alpha) \)};
\node[output, xshift=5cm, yshift=0cm](out2){90 bars \\ \( \beta \) };
\node[output, xshift=5cm, yshift=-1cm](out3){60 bars \\ \( {\mathsf{h}}(h^A_2,\beta,48,{\mathsf{p}}(P,\beta,T+48,\alpha)) \) };

\draw[arrow] (tx.east) -- (out1.west);
\draw[arrow] (tx.east) -- (out2.west);
\draw[arrow] (tx.east) -- (out3.west);

  \end{tikzpicture}
  \end{center}
  \caption{Alice's commitment transaction $\tau^A_2$ with the bet that Bob will not have a proof of $P$ before block height $T$}\label{fig:alicebetcomm}
\end{figure}
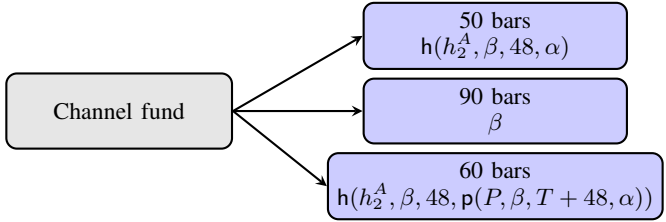

Suppose Bob has a proof of $P$ before block height $T$
and shares this proof with Alice.
Alice should agree he has won the bet and update the channel
to reflect a balance of $50$ bars for Alice and $150$ bars for Bob.
Otherwise, Bob should publish $\tau^B_2$ and publish the proof of $P$ in a document
on the Proofgold blockchain. After $48$ blocks (which is presumed to still be
before block height $T+48$), Bob can spend the outputs controlled by
${\mathsf{h}}(h^B_2,\alpha,48,\beta)$
and ${\mathsf{h}}(h^B_2,\alpha,48,{\mathsf{p}}(P,\beta,T+48,\alpha))$,
giving him $150$ bars, as desired.
Note that if Bob delays and $\tau^B_2$ is confirmed at block height $T$ or after,
then Alice will also be able to spend the 60 bars controlled by
${\mathsf{h}}(h^B_2,\alpha,{\mathsf{p}}(P,\beta,T+48,\alpha))$
and it will essentially be a competition to spend the output first.
Thus Bob should really supply the proof of $P$ at most a few blocks before
height $T$ to prevent this possibility.

Next suppose block height $T$ has arrived and Bob has not shared a proof of $P$
with Alice.
Bob should agree she has won the bet and update the channel
to reflect a balance of $110$ bars for Alice and $90$ bars for Bob.
Otherwise, Alice should publish $\tau^A_2$ at block height $T$.
After $48$ blocks, the block height will be beyond $T+48$
and Alice can spend the outputs controlled by
${\mathsf{h}}(h^A_2,\beta,48,\alpha)$
and
${\mathsf{h}}(h^A_2,\beta,{\mathsf{p}}(P,\beta,T+48,\alpha))$,
giving her 110 bars, as desired.
However, Bob still has the possibility of finding and publishing a proof of $P$
on chain before the 48 confirmations have passed.
In this case, after 48 confirmations both Alice and Bob will
race to spend the 60 bars controlled by
${\mathsf{h}}(h^A_2,\beta,{\mathsf{p}}(P,\beta,T+48,\alpha))$.

Given this behavior, it is more accurate to say Bob is betting he will have
a proof of $P$ at the latest a few blocks before $T$,
and Alice is betting Bob will not have a proof of $P$ by block $T+N$
(e.g., $T+48$ in our example).
If Bob obtains a proof during the uncertain period from $T$ to $T+N$,
either could be considered the winner of the bet.
In such a case, it would arguably be in Bob's interest to withhold
sharing the proof and make a new bet before revealing it -- as he
would be certain of winning the new bet and have only a partial
chance of winning the old bet.

The appropriate commands to support payment channels with
proofs have been implemented in the Proofgold Lava client~\cite{Proofgold2022}
and are shortly presented in Section~\ref{s:impl}.

\section{A Lightning Network with Proofs}\label{s:goldlightning}

Similar to the way payment channels can be combined to form the lightning network,
it seems possible to do the same with these ``bets'' about the existence of proofs
by a time limit.
In simple terms, Alice could bet Bob as above (50+10=60 bars) that there will not be a proof of $P$ by Block $T$.
Bob could later hedge by betting Charlie 50+10=60 bars there will not be a proof of $P$ by Block $T$.
In this case, if no one proves $P$ by Block $T$, Alice will gain 10 bars, Bob's total balance
will not change (losing 10 bars on the first channel but gaining 10 bars on the second channel)
and Charlie will lose 10 bars.
On the other hand, if someone proves $P$ by Block $T$, Alice will lose 50 bars,
Bob's total balance will not change (gaining 50 bars on the first channel but losing 50 bars on the second channel)
and Charlie will gain 50 bars.
Effectively, Bob will have sold his side of the bet to Charlie.

In the case of three actors, it is useful to consider the motivations
of each actor. If Charlie finds a proof of $P$, he is motivated
to share the proof with Bob to win one bet
and then Bob is motivated to share the proof with Alice to win the other bet.
In general, there is a motivation for proofs to propagate from those betting
there will be a proof to those betting there will not be a proof.
On the other hand, if Alice finds a proof of $P$, she is not motivated to share it at all.
As long as Charlie does not have access to a proof of $P$ by the time limit,
he has technically lost the bet (even if Alice has obtained a proof of $P$).
This suggests a third possibility that both Alice and Charlie lose the bet.
Suppose Bob finds a proof of $P$.
He is motivated to share the proof with Alice,
but not to share the proof with Charlie so that he can potentially win both bets.
If Alice wishes to protect against such a possibility, then
she could agree that whenever her channel counterparty (Bob) sends
her a proof of $P$, she makes that proof publicly available immediately.
In this case, Charlie will have access to the proof as well.
Alice will lose nothing by making the proof public.
Alice will also lose nothing by allowing Bob (who was not intended to
be a party to the bet) to win both bets,
so the appropriate behavior when a middleman finds the proof
is admittedly questionable.

\section{Prediction Market}\label{s:pmarket}

Suppose again that
Alice is willing to bet 50 bars Bob will have no proof of $P$ by Block $T$
and Bob is willing to bet 10 bars he will have a proof of $P$ by Block $T$.
This can be interpreted as the two parties agreeing that the probability
that Bob will have a proof of $P$ by Block $T$
is $0.1667$ (1 chance out of 6).
Alice is incentivized to take the highest counteroffer (since she will then
maximize her payout if no proof appears in time).
If Charlie is willing to bet Alice 50 bars he will have a proof of $P$ by Block $T$,
then Alice will prefer to make the bet with Charlie than Bob.
The bet with Charlie can be interpreted as the probability
that Charlie will have a proof of $P$ by Block $T$ is $0.5$.
In an extreme case, before Alice has accepted a counteroffer,
Daria may secretly find a proof of $P$. It is then in Daria's
interest to bet Alice as much as possible (since Daria is sure
she will win the bet). Thus when someone (other than Alice) has a proof, the
bets will reflect that by making the apparent probability
approach $1$. If Alice has a proof, she will presumably no longer offer
to make such bets.

On the other hand, suppose Alice believes $P$ is not provable (e.g.,
she has privately proven its negation).
She can freely make arbitrary bets that $P$ will not be proven by a deadline
as she will win every bet (unless, of course, the relevant theory turns out to be inconsistent).
Over time, it seems likely there will be fewer counterparties willing
to bet Alice, even with very low contributions to the bet.
This would push the apparent probability towards $0$.

It is likely the odds of the bet will change over time based on
the perceived likelihood that $P$ will be proven by Block $T$.
This change would be reflected in multiple participants making new bets
which would give the lightning network with proofs many properties of a prediction market.
A similar idea for a prediction market regarding the potential truth
of mathematical statements was described in~\cite{Su18}.

\section{Implementation and Experiments}\label{s:impl}

We have added and evaluated payment channels operations in Proofgold
using multisig addresses and hashed timelock contracts. The key
operations available for payment channels include:

\begin{itemize}
\item Creating a channel: Using \textsc{createchannel} to set up a 2-of-2
  multisig address and generate initial commitment transactions.

\item Updating channel state: Creating new commitment transactions
  with updated balances and exchanging partially signed transactions
  between parties.

\item Closing a channel: Either cooperatively by both parties signing
  a closing transaction, or unilaterally by publishing the latest
  commitment transaction.

\item Adding conditional payments: Using HTLCs to create outputs that
  can be spent based on revealing a secret or after a timeout.

\item Adding bet-like conditional payments: Using a combination of
  HTLCs and prop timelock contracts (PTLCs) to create outputs that
  depend on whether a proposition is proven by a certain block height.
\end{itemize}

The implementation has been tested with various scenarios, in particular
considering various attacks on the protocol. The particular tests include
opening channels, updating balances, adding conditional payments, and
closing channels both cooperatively and unilaterally\footnote{\texttt{https://github.com/ckaliszyk/proofgold-lava/blob/\\69fc329ab1175ab01e69385abb88402a4ee447e4/doc/\\paymentchannels-pfg.md}}. The tests cover corner cases like trying to
use revoked commitment transactions and racing to claim HTLC outputs.
We also show the bet-like conditional payments using PTLCs, that can
allow parties to wager on whether certain propositions will be proven
within a given time.

\section{Related Work}

Decker and Wattenhofer~\cite{DeckerW15} have proposed an alternative
fast payment network for Bitcoin. It remains to be seen if a similar approach
can also be extended to payment channels with proofs.
Miller et al~\cite{Miller19} discuss approaches to reduce the worst case
of the number of hops in a Layer 2 network on top of any blockchain based
currency.
A more thorough analysis of the security of the proposed lightning network
for proofs, similar to that done for the Bitcoin lightning network
also remains open~\cite{KapposYPKDMM21}.
Naor and Keidar consider payment channels in asynchronous money
transfer systems~\cite{NaorK22}.
Dziembowski et al.~\cite{DziembowskiFH18} introduced general state channel networks which
generalize payment networks.

\section{Conclusion}

We have described an extension of payment channels in Proofgold
to allow counterparties to bet on whether a certain proposition
will be proven by a certain time. The payment channels with proofs
have been implemented in the Proofgold Lava repository version
(\texttt{\small https://github.com/ckaliszyk/proofgold-lava/}) and have been tested.

These payment channels can be combined into a lightning network
allowing parties to bet on propositions via combining bets
through a route consisting individual channels. This would require
handling the complex challenges of routing and liquidity management
within a full lightning network, in particular we would need to investigate
dealing with conditional ``proof bets'' across multiple channels.

We have so far provided the necessary foundational extensions;
however, as this is developed into a complete framework, practical
performance metrics or simulations could be conducted to assess the
scalability or latency of the protocol.

In the proposed setting, we have multiple proof producers and consumers
that interact. As this is the first time such a setting is considered,
it would be interesting to formalize the economic model and incentive
alignment, ensuring that proof buyers, producers, and bettors are clearly
defined. Such a model would also specify their incentives and ensure
that they are compatible.

Further future work includes more user-friendly commands for the
lightning network. Furthermore, the bets can be interpreted as
indicating a probability, opening up the use of the network as a
prediction market for the provability of mathematical statements by a
given deadline.

\section*{Acknowledgment}

We want to thank Thibault Gauthier who provided some of the important ideas ultimately leading to this work.



\begin{thebibliography}{1}
\providecommand{\url}[1]{#1}
\csname url@samestyle\endcsname
\providecommand{\newblock}{\relax}
\providecommand{\bibinfo}[2]{#2}
\providecommand{\BIBentrySTDinterwordspacing}{\spaceskip=0pt\relax}
\providecommand{\BIBentryALTinterwordstretchfactor}{4}
\providecommand{\BIBentryALTinterwordspacing}{\spaceskip=\fontdimen2\font plus
\BIBentryALTinterwordstretchfactor\fontdimen3\font minus
  \fontdimen4\font\relax}
\providecommand{\BIBforeignlanguage}[2]{{%
\expandafter\ifx\csname l@#1\endcsname\relax
\typeout{** WARNING: IEEEtran.bst: No hyphenation pattern has been}%
\typeout{** loaded for the language `#1'. Using the pattern for}%
\typeout{** the default language instead.}%
\else
\language=\csname l@#1\endcsname
\fi
#2}}
\providecommand{\BIBdecl}{\relax}
\BIBdecl

\bibitem{LightningNetworkWhitePaper}
J.~Poon and T.~Dryja, ``The bitcoin lightning network: Scalable off-chain
  instant payments,'' 2016,
  https://lightning.network/lightning-network-paper.pdf.

\bibitem{Proofgold2022}
\BIBentryALTinterwordspacing
C.~E. Brown, C.~Kaliszyk, T.~Gauthier, and J.~Urban, ``Proofgold: Blockchain
  for formal methods,'' in \emph{4th International Workshop on Formal Methods
  for Blockchains, FMBC@CAV 2022, August 11, 2022, Haifa, Israel}, ser. OASIcs,
  Z.~Dargaye and C.~Schneidewind, Eds., vol. 105.\hskip 1em plus 0.5em minus
  0.4em\relax Schloss Dagstuhl - Leibniz-Zentrum f{\"{u}}r Informatik, 2022,
  pp. 4:1--4:15. [Online]. Available:
  \url{https://doi.org/10.4230/OASIcs.FMBC.2022.4}
\BIBentrySTDinterwordspacing

\bibitem{BrownPak19}
C.~E. Brown and K.~P\k{a}k, ``A tale of two set theories,'' in
  \emph{Intelligent Computer Mathematics - 12th International Conference,
  {CICM} 2019, Prague, Czech Republic, July 8-12, 2019, Proceedings}, ser.
  Lecture Notes in Computer Science, C.~Kaliszyk, E.~C. Brady, A.~Kohlhase, and
  C.~S. Coen, Eds., vol. 11617.\hskip 1em plus 0.5em minus 0.4em\relax
  Springer, 2019, pp. 44--60.

\bibitem{Su18}
B.~Su, ``Mathcoin: {A} blockchain proposal that helps verify mathematical
  theorems in public,'' \emph{{IACR} Cryptol. ePrint Arch.}, vol. 2018, p. 271,
  2018.

\bibitem{DeckerW15}
\BIBentryALTinterwordspacing
C.~Decker and R.~Wattenhofer, ``A fast and scalable payment network with
  bitcoin duplex micropayment channels,'' in \emph{Stabilization, Safety, and
  Security of Distributed Systems - 17th International Symposium, {SSS} 2015,
  Edmonton, AB, Canada, August 18-21, 2015, Proceedings}, ser. Lecture Notes in
  Computer Science, A.~Pelc and A.~A. Schwarzmann, Eds., vol. 9212.\hskip 1em
  plus 0.5em minus 0.4em\relax Springer, 2015, pp. 3--18. [Online]. Available:
  \url{https://doi.org/10.1007/978-3-319-21741-3\_1}
\BIBentrySTDinterwordspacing

\bibitem{Miller19}
\BIBentryALTinterwordspacing
A.~Miller, I.~Bentov, S.~Bakshi, R.~Kumaresan, and P.~McCorry, ``Sprites and
  state channels: Payment networks that go faster than lightning,'' in
  \emph{Financial Cryptography and Data Security - 23rd International
  Conference, {FC} 2019, Frigate Bay, St. Kitts and Nevis, February 18-22,
  2019, Revised Selected Papers}, ser. Lecture Notes in Computer Science,
  I.~Goldberg and T.~Moore, Eds., vol. 11598.\hskip 1em plus 0.5em minus
  0.4em\relax Springer, 2019, pp. 508--526. [Online]. Available:
  \url{https://doi.org/10.1007/978-3-030-32101-7\_30}
\BIBentrySTDinterwordspacing

\bibitem{KapposYPKDMM21}
\BIBentryALTinterwordspacing
G.~Kappos, H.~Yousaf, A.~M. Piotrowska, S.~Kanjalkar, S.~Delgado{-}Segura,
  A.~Miller, and S.~Meiklejohn, ``An empirical analysis of privacy in the
  lightning network,'' in \emph{Financial Cryptography and Data Security - 25th
  International Conference, {FC} 2021, Virtual Event, March 1-5, 2021, Revised
  Selected Papers, Part {I}}, ser. Lecture Notes in Computer Science,
  N.~Borisov and C.~D{\'{\i}}az, Eds., vol. 12674.\hskip 1em plus 0.5em minus
  0.4em\relax Springer, 2021, pp. 167--186. [Online]. Available:
  \url{https://doi.org/10.1007/978-3-662-64322-8\_8}
\BIBentrySTDinterwordspacing

\bibitem{NaorK22}
\BIBentryALTinterwordspacing
O.~Naor and I.~Keidar, ``On payment channels in asynchronous money transfer
  systems,'' in \emph{36th International Symposium on Distributed Computing,
  {DISC} 2022, October 25-27, 2022, Augusta, Georgia, {USA}}, ser. LIPIcs,
  C.~Scheideler, Ed., vol. 246.\hskip 1em plus 0.5em minus 0.4em\relax Schloss
  Dagstuhl - Leibniz-Zentrum f{\"{u}}r Informatik, 2022, pp. 29:1--29:20.
  [Online]. Available: \url{https://doi.org/10.4230/LIPIcs.DISC.2022.29}
\BIBentrySTDinterwordspacing

\bibitem{DziembowskiFH18}
\BIBentryALTinterwordspacing
S.~Dziembowski, S.~Faust, and K.~Host{\'{a}}kov{\'{a}}, ``Foundations of state
  channel networks,'' \emph{{IACR} Cryptol. ePrint Arch.}, p. 320, 2018.
  [Online]. Available: \url{https://eprint.iacr.org/2018/320}
\BIBentrySTDinterwordspacing

\end{thebibliography}
\end{document}